\documentclass[10pt,a4paper]{article}
\usepackage[utf8]{inputenc}
\usepackage[english]{babel}
\usepackage{amsmath}
\usepackage{amsfonts}
\usepackage{amssymb}
\usepackage{graphicx}

\usepackage{amsthm,amsmath,amsfonts,amssymb}
\usepackage[authoryear]{natbib}
\usepackage[utf8]{inputenc}
\usepackage[english]{babel}
\usepackage{amsmath}
\usepackage{amsfonts}
\usepackage{amssymb, amsthm}
\usepackage{graphicx}
\usepackage{color}
\usepackage{fullpage}
\usepackage{amssymb,amsmath,dsfont}
\usepackage{natbib}
\usepackage{csquotes}
 \usepackage{tabularx}
\newcommand{\bw}{\boldsymbol{w}}

\newcommand{\tz}{\textbf{z}}

\newcommand{\bX}{\boldsymbol{X}}
\newcommand{\bZ}{\boldsymbol{Z}}
\newcommand{\bz}{\boldsymbol{z}}

\newcommand{\tX}{\textbf{X}}

\newcommand{\ty}{\textbf{y}}
\newcommand{\bY}{\boldsymbol{Y}}
\newcommand{\bW}{\boldsymbol{W}}
\newcommand{\bV}{\boldsymbol{V}}

\newcommand{\bu}{\boldsymbol{u}}

\newcommand{\bv}{\boldsymbol{v}}

\newcommand{\by}{\boldsymbol{y}}
\newcommand{\bvarepsilon}{\boldsymbol{\varepsilon}}
\newcommand{\bbeta}{\boldsymbol{\beta}}
\newcommand{\brho}{\boldsymbol{\rho}}
\newcommand{\bnu}{\boldsymbol{\nu}}
\newcommand{\bPi}{\boldsymbol{\Pi}}

\newcommand{\bpsi}{\boldsymbol{\psi}}

\newcommand{\btheta}{\boldsymbol{\theta}}

\newcommand{\bm}{\boldsymbol{m}}
\newcommand{\balpha}{\boldsymbol{\alpha}}

\newcommand{\bmu}{\boldsymbol{\mu}}

\newcommand{\lp}{\ell_{\text{pen}}}
\newcommand{\bOm}{\boldsymbol{\Omega}}
\newcommand{\bth}{\boldsymbol{\theta}}
\newcommand{\bmm}{\boldsymbol{m}}
\newcommand{\mhat}{\widehat{\boldsymbol{m}}}

\DeclareMathOperator*{\argmax}{arg\,max}

\newtheorem{lemma}{Lemma}
 \newcommand{\footremember}[2]{%
    \footnote{#2}
    \newcounter{#1}
    \setcounter{#1}{\value{footnote}}%
}
\newcommand{\footrecall}[1]{%
    \footnotemark[\value{#1}]%
} 
\title{Investigating swimming technical skills by a double partition clustering of multivariate functional data allowing for dimension selection}
\author{Antoine Bouvet\footremember{ENS}{Univ. Rennes, ENS Rennes, M2S Laboratory-EA 7470}\footremember{Inria}{Inria Rennes Bretagne Atlantique, MIMETIC}\footremember{CREST}{Univ. Rennes, Ensai, CNRS, CREST—UMR 9194 }, Salima El Kolei\footrecall{CREST} and Matthieu Marbac\footrecall{CREST}}
\begin{document}
\maketitle
 
%\title{A sample article title with some additional note\thanksref{T1}}
 %\thankstext{T1}{A sample of additional note to the title.}
 
%%%%%%%%%%%%%%%%%%%%%%%%%%%%%%%%%%%%%%%%%%%%%%%
%% Only one address is permitted per author. %%
%% Only division, organization and e-mail is %%
%% included in the address.                  %%
%% Additional information can be included in %%
%% the Acknowledgments section if necessary. %%
%% ORCID can be inserted by command:         %%
%% \orcid{0000-0000-0000-0000}               %%
%%%%%%%%%%%%%%%%%%%%%%%%%%%%%%%%%%%%%%%%%%%%%%%

\begin{abstract}
Investigating technical skills of swimmers is a challenge for performance improvement, that can be achieved by analyzing multivariate functional data recorded by Inertial Measurement Units (IMU). To investigate technical levels of front-crawl swimmers, a new model-based approach is introduced to obtain two complementary partitions reflecting, for each swimmer, its swimming pattern and its ability to reproduce it. Contrary to the usual approaches for functional data clustering, the proposed approach also considers the information of the residuals resulting from the functional basis decomposition. Indeed, after decomposing  into functional basis  both the original signal (measuring the swimming pattern) and the signal of squared residuals (measuring the ability to reproduce the swimming pattern), the method fits the joint distribution of the coefficients related to both decompositions by considering dependency between both partitions. Modeling this dependency is mandatory since the difficulty of reproducing a swimming pattern depends on its shape. Moreover, a sparse decomposition of the distribution within components that permits a selection of the relevant dimensions during clustering is proposed. The partitions obtained on the IMU data aggregate the kinematical stroke variability linked to swimming technical skills and allow relevant biomechanical strategy for front-crawl sprint performance to be identified.
\end{abstract}

%%%%%%%%%%%%%%%%%%%%%%%%%%%%%%%%%%%%%%%%%%%%%%
%% Please use \tableofcontents for articles %%
%% with 50 pages and more                   %%
%%%%%%%%%%%%%%%%%%%%%%%%%%%%%%%%%%%%%%%%%%%%%%
%\tableofcontents

%%%%%%%%%%%%%%%%%%%%%%%%%%%%%%%%%%%%%%%%%%%%%%
%%%% Main text entry area:
\section{Introduction}
Tracking of technical skills by comprehensive training monitoring is a main challenge for sport performance improvment \citep{bompa2018periodization}, that becomes central for swimming since such a sport requires efficient movement through high levels of motor abilities that are not easily observed by the coaches due to the aquatic environment. Swimming is a continuous sequence of periodic, coordinated and synchronous body movements \citep{maglischo2003}. It defines the technique by the repetition of similar but not identical stroke patterns constituted of instabilities called biomechanical variability (\cite{fernandes2022velocity}; \cite{preatoni2013movement}). Factors concerning stroking, such as the variability of swimming patterns play a major role in generating swimming speed \citep{figueiredo2013} because they are related to swimming efficiency (\cite{figueiredo2012intracycle}; \cite{ribeiro2013biomechanical}) and differ according to swimming performance levels (\cite{matsuda2014intracyclic}; \cite{seifert2016behavioural}). Thus, technical skills can be described by this biomechanical variability and quantified through two complementary and associated aspects of motion: the swimming pattern and the ability to reproduce it \citep{fernandes2022intra}. Indeed, development of automatic methodologies supporting investigation of both of these components, complementing the observations of coaches, is promising. To do this, Inertial Measurement Units (IMU) are used to provide embedded kinematical data collection regarding sports related movements about through tri-axial accelerometer and gyroscopic temporal records \citep{camomilla2018}. They have become more readily available in swimming and provide on-board technical assessments to be made \citep{guignard2017}. However, the current literature is focused on descriptions of average swimming patterns justifying qualitative analysis (\cite{staniak2016}; \cite{staniak2018}), performance group comparison based on mean stroke patterns \citep{ganzevles2019} or multi-cycle variability assessment through stroke descriptors \citep{dadashi2016front}. On the one hand, some of these studies are focused on accelerometric norms which results in masking some kinematical dimensions that can be informative regarding technical skills. On the other hand, gyroscopic assessments of IMUs are rarely included whereas rotation motions are determinant in the swimming technique \citep{maglischo2003}. Lastly, all of these previous works were restricted to empirical indicators describing stroke patterns without statistical modelling of IMU data taking their functional nature into account. Hence, thus leads to limited insights regarding underlying kinematical variability defining technical skills and making their conclusions weak or not representative. Their main limitation is due to the absence of a powerful statistical model able to analyse the complex multivariate dynamics of swimming patterns.

Statistical analysis of IMU data supporting monitoring of kinematical variability can be achieved by investigating disparities between swimmers with respect to their technical skills. This could be achieved by a dependent double partition clustering on multivariate functional periodic IMU data measuring the swimming pattern (\emph{i.e.,} first latent partition) as well as the ability to reproduce it (\emph{i.e.,} second latent partition). Indeed, the type of stroke pattern directly impacts its repeatability since high biomechanical variability is for example more difficult to maintain between cycles compared to low levels of stroke variability. Moreover, dimension selection performed during this double partition clustering could allow the IMU axes to be identified  for explaining disparities between swimmers and associated technical skills. This double partition clustering presents new prospects for sports science including, on the one hand a deeper understanding of stroke patterns and subsequent technical skills under real condition of swimming training, and on the other hand, raising the comprehension of relevant technical strategies during swimming by taking into account  the individual stroke profiles of swimmers.

Different statistical methods allow functional data to be clustered (see the review of \cite{jacques2014functional}). Traditionally,
they consider a decomposition of the functional data  into a functional basis  \citep{ray2006functional}, and then use classical multivariate methods for finite dimensional data directly on these basis coefficients \citep{abraham2003unsupervised}. Considering this approach, different model-based clustering methods have been developed for univariate functional data (\cite{bouveyron2011model}; \cite{bouveyron2015discriminative}; \cite{bouveyron2019model}) as well as in the multivariate case (\cite{ieva2011multivariate}; \cite{yamamoto2012clustering}; \cite{jacques2013funclust}; \cite{yamamoto2014functional}; \cite{yamamoto2017dimension}; \cite{schmutz2020clustering}). Multivariate approaches working with a set of several time series describing the same individual are relevant for deeper data analysis \citep{bouveyron2019model}. In sports science, model-based clustering approaches for functional data provide a useful framework for new insights on performance (\cite{leroy2018}; \cite{leroy2020multi}) that can outpace classical approaches relying on empirical analysis that lead to limited kinematical information \citep{mallor2010changes}. In this way, the literature includes motion pattern analysis and description of different performance groups in some cyclic sports such as running and cycling (\cite{liebl2014ankle}; \cite{forrester2015effect}; \cite{helwig2016smoothing}). Note that, only the work of \cite{schmutz2018donnees} uses multivariate functional clustering with a model-based approach on IMU data, for sports application. Their results outperfom classical biomechanical and physical models in the literature suggesting that this framework is relevant for new insights on kinematics. For all the methods considering functional basis decompositions, the loss of information due to the approximation of the original data into a functional basis is neglected. In the context of investigating technical swimming skills, these methods could only provide clusters of swimmers having the same swimming pattern, but these clusters can be composed by swimmers having different abilities to reproduce it. Indeed, the ability of reproducing the swimming pattern corresponds to the dispersion around it and this information is lost by using the basis decomposition. Thus, for investigating swimming techniques, it is also crucial to keep the information of the dispersion around the swimming pattern. 

The objective of estimating two partitions from the same data set has some connections with different statistical approaches. Note that our objective is different from the co-clustering or bi-clustering objective. Indeed co-clustering \citep{slimen2018model,bouveyron2022co} aims to estimate one partition among the rows (the observations) and one partition among the columns (the variables) while our objective is to find two partitions among the observations. Therefore, our objective is closer to the objective of multi-partitions clustering \citep{GALIMBERTI2007520,Galimberti2017,marbac2019tractable}  but two important differences should be pointed out. First, the approach considering multiple partitions tries to estimate the subsets of the variables that are related to the different latent variables. In our context, these subsets of variables are known since they are naturally defined by the meaning of the latent variables. Indeed, the variables that are related to the decomposition of the observed data into functional basis provide  information on the swimming pattern while the variables related to the dispersion of the residuals resulting to this basis decomposition provides information on the ability to reproduce the swimming pattern.  Second, most of the approaches considering multiple partitions need to assume independence between the latent variables for computational reasons (see for instance \citet{marbac2019tractable} and \citet{Galimberti2017} for some relaxations of this assumption) while, in our context modeling these dependencies is a great of interest. Indeed, it seems to be natural to consider that the more complex the swimming pattern is, the more difficult its reproductibility should be.

Some model-based clustering approaches have been developed to perform a selection of the variables \citep{tadesse2005bayesian,raftery2006variable,marbac2017variable}. The main idea is to consider that only a small subset of the variables explain the true underlying clusters.  Thus, selecting variables is very challenging in clustering because the role of a variable (relevant or irrelevant for clustering) is defined from a variable that is not observed. Thus, the selection of the variables and the clustering need to be performed simultaneously. Note that selecting the variables in clustering has two strong benefits.  First, it facilitates the interpretation of the different components as it only has to be done on the subset of discriminative variables.  Second, it improves the accuracy of  the  estimators  because  it  reduces  the  number  of  estimators  to  be  considered. In the context of clustering multivariate functional data, to the best of our knowledge, no methods can lead to a detection of the dimensions that are relevant for clustering. However, one can consider a natural extension of the model-based clustering approach performing feature selection. Indeed, when the functional data are decomposed into a functional basis, claiming that a dimension of the functional data is not relevant for clustering means that all the coefficients of the functional basis, related to this dimension, are not informative for the clustering.

In this article, we propose a double partition clustering approach to investigate technical swimming skills using IMU data that are periodic due to the repetition of a cyclic stroke pattern structure (\cite{Mooney2015} ; \cite{silva2011wearable}). Moreover, the proposed model-based approach allows for the identification of the discriminative dimensions for each partition. After decomposing both of the original signals (\emph{i.e.,} measuring the swimming pattern) and the signals of squared residuals (\emph{i.e.,} measuring the ability to reproduce the swimming pattern) into a Fourier basis, the method fits the joint distribution of the coefficients related to both decompositions by considering dependency between both partitions. Modelling this dependency is important because the difficulty of reproducing a swimming pattern depends on its shape and then belongs to technical skills. The model considers that the information about the swimming pattern that measures the kinematical smoothness linked to the energy economy of the technique (\emph{i.e.,} the first partition) is contained in the coefficients arising from the decomposition of the original signal while the information about the ability of reproducing the pattern (\emph{i.e.,} the second partition) is contained in the coefficients arising from the decomposition of the squared noise signal. As usual for a standard model-based approach to cluster functional data, a  sparse decomposition of the distribution within components is used. Here, we consider a conditional independence between the coefficients of different dimensions within the component. This assumption allows an automatic selection to be made of the relevant dimensions during clustering that improves the accuracy of the estimates and that highlights the dimensions (\emph{i.e.,} IMU axis) that are discriminative for the swimming pattern and for the ability to reproduce this pattern.

Interpretation of this double partition clustering is done by computing some biomechanical parameters from the literature on the reconstructed mean curves from the first clustering (\emph{i.e.,} assessing the swimming pattern) and by analyzing the gathered partitions regarding swimming speed. Finally, a combination of both clusterings can be used to model kinematical variability and allows technical skills of swimmers to be investigated. This is the first method that permits analysis of IMU data for swimming and provides an easy-to-use statistical tool for practitioners. Notably, it may assist on-field coaches observations for providing adapted and effective feedback based on \emph{in-situ} individual technical profiling of their swimmers during daily practice.

The paper is organized as follows. 
Section~\ref{sec:data} introduces the context of the study and the SWIMU data. 
Section~\ref{sec:method} presents the double clustering model-based framework and dimension selection for multivariate functional data. Identifiability issues of this new model are investigated and presented. Section~\ref{sec:est} is devoted to model inference. A numerical study of this approach is performed on simulated data in Section~\ref{sec:simulation}.
Section~\ref{sec:appli} is devoted to the analysis of the SWIMU data and the consequent biomechanical interpretation for technical swimming skills analysis.
Some concluding remarks and extension of this work are finally discussed in Section~\ref{sec:conclusions}.

\section{SWIMU data description}\label{sec:data}

\begin{figure}[ht!p]
    \centering
    \includegraphics[scale=0.5]{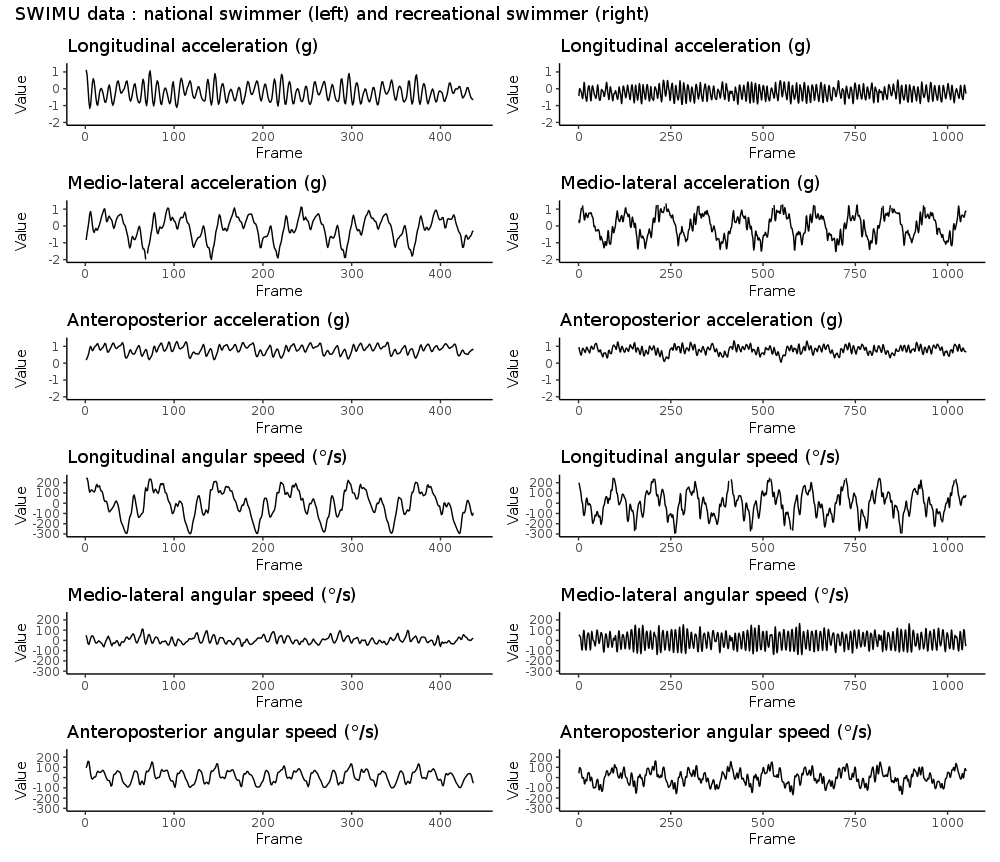}
    \caption{Data of two subjects (national level on left and recreational level on right) of the SWIMU dataset for one front-crawl lap performed at maximal intensity}
    \label{plotmethodo}
\end{figure}

The database used in this study gathered part of the data from several sport-science experimental sessions in a swimming pool. It includes \textit{n} = 68 front-crawl pool length performances at maximal intensity by participants ranging from recreational to international level. All athletes signed an informed consent form in agreement with the French ethical committee (approval obtained under the reference: 2021-A00250-41). Each swimmer $i$ was observed on $[0,T_i]$ during one pool length.\\
The participants were instrumented with one waterproofed IMU (Xsens DOT, Xsens Technologies B.V, Enschede, The Netherlands) located on the sacrum. Note that, according to the literature, this is the best anatomical positioning for mono-sensor measurements \citep{Rad2020novel}. The IMU was attached with double-sided tape and secured with waterproof medical adhesive (Tegarderm, 3M, Cergy-Pontoise, France). The sensor was composed of a 3D accelerometer and a 3D gyroscope sampling at the same frequency of 60 Hz with a full scale sets at respectively $\pm$8g and $\pm$1000°.s, respectively. The IMU described a coordinate system defined as x-axis pointing cranially, y-axis pointing laterally and z-axis pointy posteriorly. \\
Raw sensor data was filtered using a second order Butterworth low-pass filter with a 10Hz cut-off frequency and downsampled to 50Hz in order to match the data with the human activity recognition algorithm   \citep{delhaye2022}. We use it in order to keep only the front-crawl phase during the lap and to remove all the others phases including for example the wall-push, underwater and turn or end phase. To avoid a potential bias of this segmentation and kinematic disturbance during the first strokes after swimmer's breakout (\emph{i.e.,} transition between underwater and front-crawl phases) and the last stroke before the swimmer's turn or finish (\emph{i.e.,} transition between front-crawl phases and turn or rest phases), we removed 1s before and after the first and the last front-crawl frame classified by the algorithm. Figure \ref{plotmethodo} gives an example of SWIMU dataset for 2 subjects of different performance level. \\

\section{Model-based approach for estimating a double partition from multivariate functional data}\label{sec:method}
This section presents the new model-based approach for estimating a double partition from multivariate functional data that allows for a selection of the relevant dimensions. The first part presents the general approach for estimating a double partition while the second part focuses on the selection of the dimensions that are relevant for each unobserved partition.

\subsection{Mixture model on basis coefficients}
We consider a random sample $\tX=(\bX_1^\top,\ldots,\bX_n^\top)^\top$ composed of $n$ independent and identically distributed multivariate time series. Each individual $i$ is described by a $J$-dimensional discrete-time  time series $\bX_i=(\bX_{i1}^\top,\ldots,\bX_{iJ}^\top)^\top$, where $\bX_{ij}=(X_{ij}(1),\ldots,X_{ij}(T_i))^\top$ and $X_{ij}(t)\in \mathbb{R}$ denotes the value of dimension $j$ of the multivariate time series measured on subject $i$ at time $t$, $T_i$ being the length of the multivariate time series recorded on subject $i$.

Each univariate time series admits a basis expansion (see \citet{ramsey2005functional}) leading to
\begin{equation}\label{mod}
    X_{ij}(t) = \bY_{ij}^\top \bpsi_{j}(t) + \varepsilon_{ij}(t),
\end{equation}
where $\bY_{ij}\in\mathbb{R}^{G_j}$ groups the $G_j$ basis coefficients, $\bpsi_{j}(t)=(\psi_{j1}(t),\ldots,\psi_{jG_j}(t))^\top$ is the vector containing the values of the $G_j$ basis functions evaluated at time $t$ and where the random noise $\varepsilon_{ij}(t)$ is supposed to be centered given the natural filtration $\mathcal{F}_{i}(t)$ (\emph{i.e.,} $\mathbb{E}[\varepsilon_{ij}(t) \mid \mathcal{F}_i(t-1)]=0$). 
Let $\bY_i=(\bY_{i1}^\top,\ldots,\bY_{iJ}^\top)^\top\in\mathbb{R}^G$ be the vector of length $G=\sum_{j=1}^J G_j$ that gathers the basis coefficients of subject $i$ for the $J$ dimensions and let $\bvarepsilon_i=(\bvarepsilon_{i1}^\top,\ldots,\bvarepsilon_{iJ}^\top)^\top$ be the vector of the errors of individuals $i$ where $\bvarepsilon_{ij}=(\varepsilon_{ij}(1),\ldots,\varepsilon_{ij}(T_i))^\top$. In addition, we express each univariate time series of the square error term in a functional basis with $H_j$ elements as follows
\begin{equation}\label{mod2}
\varepsilon_{ij}^2(t) = \bZ_{ij}^\top \bPi_j(t) + \xi_{ij}(t),
\end{equation}
where $\bZ_{ij}\in\mathbb{R}^{H_j}$ groups the $H_j$ basis coefficients, $\bPi_{j}(t)=(\Pi_{j1}(t),\ldots,\Pi_{jH_j}(t))^\top$ is the vector containing the values of the $H_j$ basis functions evaluated at time $t$,  $\mathbb{E}[\xi_{ij}(t)\mid \mathcal{F}_i(t-1)]=0$.

The decomposition defined by \eqref{mod} and \eqref{mod2} considers a general function basis.  To model  the cyclical pattern of the swimmers' motion (see Figure \ref{plotmethodo}), it seems appropriate to use a Fourier basis for the decomposition of each time series (see \eqref{mod} and \eqref{mod2}). The period is the same among the dimensions of the functional data but the degrees of the Fourier basis can be different. Also, it is important to underline that although the choice of the basis is the same for all the swimmers it differs for each swimmer due to the different swimming period between the swimmers. This follows that, if swimmer $i$ has the same period $\beta_i$, we have $\bpsi_j(t):=\bpsi_j(t;\beta_i)$ where $\bpsi_j(t;\beta_{i}) =(\bpsi_{j1}(t;\beta_{i}),\ldots,\bpsi_{j H_j}(t;\beta_{i}))^\top$, $\bpsi_{j1}(t;\beta_{i})=1$ and $\bpsi_{j(2\ell)}(t;\beta_{i})=\cos(2\pi\ell t/\beta_{i})$ and $\bpsi_{j(2\ell+1)}(t;\beta_{i})=\sin(2\pi\ell t/\beta_{i})$ $\ell=1,\ldots,(G_j-1)/2$, and we have $\bPi_j(t):=\bPi_j(t;\beta_i)$ where $\bPi_j(t;\beta_{i}) =(\bPi_{j1}(t;\beta_{i}),\ldots,\bPi_{j H_j}(t;\beta_{i}))^\top$, $\bPi_{j1}(t;\beta_{i})=1$ and $\bPi_{j(2\ell)}(t;\beta_{i})=\cos(2\pi\ell t/\beta_{i})$ and $\bPi_{j(2\ell+1)}(t;\beta_{i})=\sin(2\pi\ell t/\beta_{i})$ $\ell=1,\ldots,(H_j-1)/2$.

 We aim at grouping the swimmers according to their swimming pattern as well as their ability to reproduce it. This implies the estimation of two latent categorical variables $\bV_i=(V_{i1},\ldots,V_{iK})^\top$ and $\bW_i=(W_{i1},\ldots,W_{1L})^\top$ that indicate the swimming pattern and the dispersion around this pattern for swimmer $i$ respectively, $K$ and $L$ denoting the number of different swimming pattern and the number of types of dispersion around this pattern. Model-based clustering of functional data is generally achieved by considering the basis coefficients as random variables whose the distribution depends on the latent variable. Since the basis coefficients $\bY_{i}$ depend on the mean behavior of $\bX_{i}$, we consider that this vector contains all the information about $\bV_i$. Similarly, since the basis coefficients $\bZ_i$ depend on the mean behavior of $\bvarepsilon_{i}$, we consider that this vector contains all the information about $\bW_i$. Thus, the model  assumes conditional independence between $\bY_i$ and $\bW_i$ given $\bV_i$ and conditional independence between $\bvarepsilon_i$ and $\bV_i$ given $\bW_i$. Finally, it allows  for dependency between $\bV_i$ and $\bW_i$ and it assumes conditional independence between $\bY_i$ and $\bvarepsilon_i$ given $\bW_i,\bV_i$. Thus, the basis coefficients follow a specific mixture model with $K\times L$ components defined by the following probability distribution function (pdf) 
 \begin{equation}\label{eq:modelgrl}
p(\by_i,\bz_i;\bmm ,\btheta)= \sum_{k=1}^K\sum_{\ell=1}^L \pi_{k\ell} f_k(\by_i;\balpha_k) g_\ell(\bz_i;\bbeta_\ell),
 \end{equation}
where $\pi_{\ell k}:=\mathbb{P}(V_{ik}=1,W_{i\ell}=1)>0$, $\sum_{k=1}^K\sum_{\ell=1}^L \pi_{k\ell}=1$, $\btheta$ groups all parameters of model $\bmm$, $f_k$ is the pdf of cluster $k$ for the swimming pattern parameterized by $\balpha_k$ that defines the conditional distribution of $\bY_i$ given $V_{ik}=1$ and $g_\ell$ is the pdf of cluster $\ell$ for the ability of reproducing the swimming pattern parameterized by $\bbeta_\ell$ that defines the conditional distribution of $\bZ_i$ given $W_{i\ell}=1$.

In this work, we consider that a cluster is composed by the subset of curves whose related coefficients arise from the same mixture components. However, some alternative definitions could be used (see for instance \citet{baudry2010combining, hennig2010methods, hennig2015true}).
 
The model defined by \eqref{eq:modelgrl} is a parsimonious mixture model that imposes equality constraints between the parameters of some pdfs of its components. It implies that the marginal distribution of $\bY_i$ is a mixture model with $K$ components and that the marginal distribution of $\bZ_i$ is a mixture model with $L$ components. Note that dependency between $\bY_i$ and $\bZ_i$ is considered by \eqref{eq:modelgrl} and thus this model is not equivalent to a product of two mixture models modeling the marginal distribution of $\bY_i$ and $\bZ_i$. During our numerical experiments, we illustrate the interest of considering the distribution of the pair $(\bY_i^\top,\bZ_i^\top)^\top$ to improve the accuracy of the estimators of the pair $(\bV_i^\top,\bW_i^\top)^\top$. The model defined by \eqref{eq:modelgrl} has connections with mixture models allowing multiple partitions to bes estimated \citep{GALIMBERTI2007520,Galimberti2017,marbac2019tractable}. However, two important differences should be pointed out. First, the approach considering multiple partitions tries to estimate the subsets of the variables that are related to the different latent variables. In our context, these subsets of variables are known since they are naturally defined by the meaning of the latent variables ($\by_i$ and $\bz_i$ define the two groups of the observed variables used to estimate the two partitions). Second, most of the approaches considering multiple partitions need to assume independence between the latent variables for computational reasons (see for instance \citet{marbac2019tractable} and \citet{Galimberti2017} for some relaxations of this assumption) while, in our context, modeling these dependencies is of great interest. Indeed, it seems to be natural to consider that the more complex the swimming pattern is, the more difficult its repeatability should be. 

Finally, we present two properties of the model. As stated by the following lemma, the identifiability of the parameters of the marginal distributions of $\bY_i$ and $\bZ_i$ leads to the identifiability of parameters $\btheta$ in the model \eqref{eq:modelgrl}.
\begin{lemma}\label{lem:idengrl}
If the parameters of the marginal distributions $\bY_i$ and $\bZ_i$ are identifiable, then the parameters of model defined by \eqref{eq:modelgrl} are identifiable.
\end{lemma}
The second property states that considering the dependency between the two latent partitions, and thus using the joint distribution of $(\bY_i^\top,\bZ_i^\top)^\top$ for estimating the two partitions leads to a better estimator of both partitions than considering an estimator of $\bV_i$ based on $\bY_i$ and an estimator of $\bW_i$ based on $\bZ_i$ when $\bV_i$ and $\bW_i$ are not independent. Let  $\Upsilon_V(\bY_i)$ and $\Upsilon_V(\bY_i,\bZ_i)$ be applications that associate an estimator of $\bV_i$ (\emph{i.e.,} a vector of length $K$ composed by zeros except for one coordinate that is equal to one) by using the \emph{MAP} rule (\emph{i.e.,} affecting an observation to the most likely cluster) based on the distribution of $\bY_i$ and $(\bY_i^\top,\bZ_i^\top)^\top$ respectively. Let $\Upsilon_W(\bZ_i)$ and $\Upsilon_W(\bY_i,\bZ_i)$ be applications that associate an estimator of $\bW_i$ (\emph{i.e.,} a vector of length $L$ composed by zeros except for one coordinate that is equal to one) by using the \emph{MAP} rule based on the distribution of $\bZ_i$ and $(\bY_i^\top,\bZ_i^\top)^\top$ respectively. The following lemma shows that the classification errors obtained by $\Upsilon_V(\bY_i)$ and $\Upsilon_W(\bZ_i)$ are strictly larger that the classification errors obtained by $\Upsilon_V(\bY_i,\bZ_i)$ and $\Upsilon_W(\bY_i,\bZ_i)$, when the latent variables are not independent.
\begin{lemma}\label{lem:partitionaccuracy}
If the model  \eqref{eq:modelgrl} holds true and if the $\bV_i$ and $\bW_i$ are not independent then, under the assumption of Lemma~\ref{lem:idengrl} and if the densities $f_k$ and $g_\ell$ are continuous for any $k$ and $\ell$, then  
$$
\mathbb{E}[\Upsilon_V(\bY_i,\bZ_i)\neq \bV_i] < \mathbb{E}[\Upsilon_V(\bY_i)\neq \bV_i] 
\text{ and }
\mathbb{E}[\Upsilon_W(\bY_i,\bZ_i)\neq \bW_i] < \mathbb{E}[\Upsilon_W(\bZ_i)\neq \bW_i] .
$$
\end{lemma}

\subsection{Parsimonious model for detecting the relevant dimensions}
Since the decompositions into the functional basis can produce high-dimensional vectors, it is usual to assume parsimonious constraints on the dependency within components (see for instance \citet{jacques2014model,schmutz2020clustering}). Here, we consider that the mixture components belong to the same parametric family and  that the coefficients related to different dimensions are conditionally independent given the latent variables. This leads to $\bY_{ij} \perp \bY_{ij'} |\bV_i$ and $\bZ_{ij} \perp \bZ_{ij'} |\bW_i$, for $j\neq j'$. We denote by $\phi_j(\cdot;\balpha_{kj})$ the $G_j$-dimensional density of $\bY_{ij}$ within component $k$ parameterized by $\balpha_{kj}$ and by $\varphi_j(\cdot;\bbeta_{\ell j})$ the $H_j$-dimensional density of $\bZ_{ij}$ within component $\ell$ parameterized by $\bbeta_{\ell j}$.  Therefore, we have
\begin{equation} \label{eq:decompogr}
f_k(\by_i;\balpha_k) = \prod_{j=1}^J \phi_j(\by_{ij}; \balpha_{kj}) \text{ and } 
g_\ell(\bz_i;\bbeta_{\ell}) = \prod_{j=1}^J \varphi_j(\bz_{ij}; \bbeta_{\ell j}). 
\end{equation}
Note that the model can consider any parametric multivariate density of $\phi_j$ and $\varphi_j$ (\emph{e.g.,} multivariate Gaussian distribution with a full or sparse covariance matrix, multivariate Student distributions,...) and thus allows the usual parametric assumptions to be made to cluster functional data based on the coefficients of their basis extension \citep{bouveyron2011model, jacques2014model}. 

The main benefits of \eqref{eq:decompogr} is that it easily permits a selection of the dimensions that are relevant for clustering in \eqref{eq:modelgrl} and so allows functional data variable selection methods for clustering to be extended. In our context, a dimension is relevant for estimating one partition if the coefficients related to this dimension do not have the same distribution among the mixture components. Therefore, dimension $j$ is irrelevant for estimating the swimming pattern if $\balpha_{1j}=\ldots=\balpha_{Kj}$ while this dimension is irrelevant for  estimating the ability of reproducing the swimming pattern if $\bbeta_{1j}=\ldots=\bbeta_{Lj}$. We denote by  $\Omega \subseteq \{1,\ldots,J\}$  and  $\Gamma \subseteq \{1,\ldots,J\}$ the indexes of the dimensions that are relevant for  estimating the swimming pattern and  for  estimating the ability of reproducing the swimming pattern respectively. Thus, for a fixed model $\bm=\{K,L,\Omega,\Gamma\}$, using \eqref{eq:modelgrl}-\eqref{eq:decompogr} and the definition of the relevant dimensions, the pdf of the observed data is defined by
\begin{multline} \label{eq:modelfinal}
f(\by_i,\bz_i;\bmm ,\btheta)=  \left[ \prod_{j\in\Omega^c} \phi_j(\by_{ij};\balpha_{1j})\prod_{j'\in\Gamma^c} \varphi_{j'}(\bz_{ij'};\bbeta_{1j'}) \right]\times \\ \sum_{k=1}^K\sum_{\ell=1}^L \pi_{k\ell}   \prod_{j\in\Omega} \phi_j(\by_{ij};\balpha_{kj}) \prod_{j'\in\Gamma} \varphi_{j'}(\bz_{ij'};\bbeta_{\ell j'})  .
\end{multline}
The following lemma gives sufficient conditions to state the identifiability of the parameters of the proposed model \eqref{eq:modelfinal}.
\begin{lemma}\label{lemma:IDfinal}
If $\text{card}(\Omega)\geq 1$, $\text{card}(\Gamma)\geq 1$, exist $j\in\Omega$ and $j'\in\Gamma$ such that the marginal distribution of $\bY_{ij}$ and $\bZ_{ij'}$ is identifiable, then the parameters of model \eqref{eq:modelfinal} are identifiable.
\end{lemma}

\section{Estimation}\label{sec:est}

\subsection{Estimation by maximizing the penalized log-likelihood}
Based on an observed sample obtained by the basis decomposition of the $n$ independent and identically distributed multivariate time-series $\ty=(\by_1^\top,\ldots,\by_n^\top)^\top$, we aim to estimate  the parameters of model  \eqref{eq:modelfinal}. The log-likelihood function is defined, for a model $\bm$, by
$$
\ell(\bth|\bmm,\ty,\tz)=\sum_{i=1}^{n} \ln p(\by_i,\bz_i;\bmm ,\btheta),
$$
and the penalized log-likelihood function is defined, for a model $\bm$ by
$$
\lp(\bth|\bmm,\ty,\tz) = \ell(\bth|\bmm,\ty,\tz) - \nu_{\bmm} c ,
$$
$c>0$ being some positive constant and  $\nu_{\bmm}$ being the dimension of the parameter space for model $\bmm$. Note that by considering $c=(\ln n)/2$, model selection is carried out according to the BIC. Considering upper-bounds on the possible number of components $K_{\max}$ and $L_{\max}$ and allowing each dimension to be relevant or not, the space of the competing models is defined by
$$
\mathcal{M}=\{\bm=\{K,L,\bOm,\Gamma\}:\, K\leq K_{\max},\, L\leq L_{\max}, \bOm\subseteq\{1,\ldots,J\}, \bOm\neq \emptyset, \Gamma\subseteq\{1,\ldots,J\}, \Gamma\neq \emptyset\}.
$$ 
We propose to perform model selection via penalized likelihood and parameter estimation by maximizing the likelihood. Thus, the estimator of the model $\mhat$ and the estimator of the parameters $\widehat{\btheta}_{\mhat}$ are defined by
\begin{equation*}
\mhat  = \argmax_{\bmm \in \mathcal{M}} \lp(\widehat{\bth}_{\bmm}|\bmm,\ty)  \text{ and } \widehat{\bth}_{\bmm}= \argmax_{\bmm \in \Theta_{\bm}}\ell(\widehat{\bth}_{\bmm}|\bmm,\ty), 
\end{equation*}
The computation of maximum likelihood estimates, for a fixed model can be achieved by an EM algorithm. Due to the large number of competing models, an exhaustive approach that consists in computing the penalized log-likelihood for each competing model is not feasible. Therefore, we propose
to extend the approach of \citet{marbac2020variable} for simultaneously estimating the subset of the relevant variables according to the penalized log-likelihood and the maximum likelihood estimates. Since this approach is an extension of a specific EM algorithm, we need to define the complete-data log-likelihood and the penalized complete-data log-likelihood. Denoting by $v_{ik}=1$ if swimmer $i$ has swimming pattern $k$ and $v_{ik}=0$ otherwise and denoting by $w_{i\ell}=1$ if swimmer $i$ has the type of dispersion $\ell$  around their swimming pattern and $w_{i\ell}=0$ otherwise, the complete-data log-likelihood being defined by
\begin{multline*}
\ell(\bth|\bmm,\ty,\tz, \bv,\bw) =\sum_{i=1}^{n}  \sum_{j\in \bOm^c}   \ln \phi_j(\by_{ij};\balpha_{1j}) + \sum_{i=1}^{n} \sum_{j\in\Gamma^c}   \ln \varphi_j(\bz_{ij};\bbeta_{1j}) + \sum_{i=1}^{n}\sum_{k=1}^K\sum_{\ell=1}^L v_{ik}w_{i\ell}\ln\pi_{k\ell} + \\
\sum_{i=1}^{n} \sum_{k=1}^K \sum_{j\in\bOm} v_{ik} \ln \phi_j(\by_{ij};\balpha_{kj}) + \sum_{i=1}^{n} \sum_{\ell=1}^L \sum_{j\in\Gamma} w_{i\ell} \ln \varphi_j(\bz_{ij};\bbeta_{\ell j}).    
\end{multline*}
The penalized complete-data log-likelihood is defined by
\begin{equation*}
\lp(\bth|\bmm,\ty,\tz, \bv,\bw)=  \ell(\bth|\bmm,\ty,\tz, \bv,\bw) - (KL-1)c - c K \sum_{j \in \bOm}  \nu_j - c \sum_{j \in \bOm^c}\nu_j- c L\sum_{j \in \Gamma}  \tau_j - c \sum_{j \in \Gamma^c}\tau_j, 
\end{equation*}
where $\nu_j$ is the dimension of parameters for one $G_j$-variate density $\phi_j$ and $\tau_j$ is the dimension of parameters for one $H_j$-variate density $\varphi_j$

\subsection{Selecting the dimensions of multivariate functional data for clustering}\label{sec:Estim}

For fixed numbers of components $(K,L)$, a specific EM algorithm can be implemented for simultaneously performing dimension selection and parameter estimation. This modified version of the EM algorithm finds the model maximizing the penalized log-likelihood for a fixed number of components. Thus, by running the algorithm for any possible pair $(K,L)$, model selection and parameter estimation can be achieved. The EM algorithm alternates between the computation of the conditional expectation of the penalized complete-data log-likelihood given the observed data, the current model and parameters and the maximization of this conditional expectation. Thus, for fixed numbers of components $(K,L)$, this algorithm starts at an initial point $\{\bmm^{[0]},\bth^{[0]}\}$ randomly sampled with $\bmm^{[0]}=\{K,L,\bOm^{[0]},\Gamma^{[0]}\}$,  and its iteration $[r]$ is composed of two steps:\\
\textbf{E-step} Computation of the fuzzy partition 
\begin{align*}
t_{ik\ell}^{[r]}:&=\mathbb{P}(V_{ik}W_{i\ell}=1\mid \bY_i,\bZ_i;\bmm^{[r-1]},\bth^{[r-1]})\\
&=\dfrac{\pi_{k\ell}^{[r-1]} \prod_{j \in \bOm^{[r-1]}} \phi_{j}(\by_{ij} ;  \balpha_{kj}^{[r-1]}) \prod_{j' \in \Gamma^{[r-1]}} \varphi_{j'}(\bz_{ij'} ;  \bbeta_{\ell j'}^{[r-1]})}{\sum_{k'=1}^K\sum_{\ell' = 1}^L \pi_{k'\ell'}^{[r-1]} \prod_{j \in \bOm^{[r-1]}} \phi_{j}(\by_{ij} ;  \balpha_{k'j}^{[r-1]}) \prod_{j' \in \Gamma^{[r-1]}} \varphi_{j'}(\bz_{ij'} ;  \bbeta_{\ell' j'}^{[r-1]})},
\end{align*}
\textbf{M-step} Maximization of the conditional expectation of the penalized complete-data log-likelihood given the observed data, the current model and parameters  over $\{\bOm, \Gamma,\bth\}$, hence
$\bmm^{[r]}=\{K,L,\bOm^{[r]},\Gamma^{[r]}\}$ with
$$\bOm^{[r]}=\left\{j:\,\kappa_j^{[r]} > 0 \right\},\Gamma^{[r]}=\left\{j:\,\lambda_j^{[r]} > 0 \right\},$$
$$ \pi_{k\ell}^{[r]}=\dfrac{n_{k\ell}^{[r]}}{n},\;
\balpha^{[r]}_{kj}=\left\{ \begin{array}{rl}
\balpha^{\star [r]}_{kj} & \text{if } j\in \bOm^{[r]} \\
\tilde{\balpha}_{kj} & \text{otherwise}
\end{array}\right. \text{ and } 
\bbeta^{[r]}_{\ell j}=\left\{ \begin{array}{rl}
\bbeta^{\star [r]}_{\ell j} & \text{if } j\in \Gamma^{[r]} \\
\tilde{\bbeta}_{\ell j} & \text{otherwise}
\end{array}\right. ,$$
where $n_{k\ell}^{[r]}=\sum_{i=1}^{n} t_{ik\ell}^{[r]}$,  $\balpha^{\star [r]}_{kj}=\argmax_{\balpha_{kj}}\sum_{i=1}^{n}  t_{ik\bullet}^{[r]} \ln \phi_j(\by_{ij};\balpha_{kj})$,  $t_{ik\bullet}^{[r]} =\sum_{\ell=1}^Lt_{ik\ell}^{[r]}$, $t_{i\bullet \ell}^{[r]} =\sum_{k=1}^K t_{ik\ell}^{[r]}$, $\tilde{\balpha}_{1j}=\argmax_{\balpha_{1j}}\sum_{i=1}^{n}  \ln \phi_j(\by_{ij};\balpha_{1j})$,  $\bbeta^{\star [r]}_{\ell j}=\argmax_{\bbeta_{\ell j}}\sum_{i=1}^{n}  t_{i\bullet \ell}^{[r]} \ln \varphi_j(\bz_{ij};\bbeta_{\ell j})$ and  where $\tilde{\bbeta}_{1j}=\argmax_{\bbeta_{1j}}\sum_{i=1}^{n}  \ln \varphi_j(\bz_{ij};\bbeta_{1j})$. The scalars $\kappa_j$ and $\lambda_j$ denote the difference between the maximum of the conditional expectation of the penalized complete-data log-likelihood given the observed data obtained when dimension $j$ is relevant and when it is irrelevant for estimating $\bV$ and $\bW$ respectively. Hence, we have
$$\kappa^{[r]}_j=\sum_{k=1}^K \sum_{i=1}^{n} t_{ik\bullet }^{[r]} \big(\ln \phi_{j}(\by_{ij} ;  \balpha^{\star [r]}_{kj})- \ln\phi_{j}(\by_{ij} ;  \tilde{\balpha}_{1j})\big) - (K-1)\nu_j c,$$
and
$$\lambda^{[r]}_j=\sum_{\ell=1}^L \sum_{i=1}^{n} t_{i\bullet \ell}^{[r]} \big(\ln \varphi_{j}(\bz_{ij} ;  \bbeta^{\star [r]}_{\ell j})- \ln\varphi_{j}(\bz_{ij} ;  \tilde{\bbeta}_{1j})\big) - (L-1)\tau_j c.$$

The resulting algorithm can be easily implemented and the parameters of distributions $\phi_j$ and $\varphi_j$ can be easily estimated by maximizing the likelihood. This algorithm keeps the property of monotonicity such that for any iteration  $[r]$, $\lp(\bth^{[r]}|\bmm^{[r]},\ty,\tz)  \geq \lp(\btheta^{[r-1]}|\bmm^{[r-1]},\ty,\tz)$.
Note that this algorithm can be implemented because of the assumption of independence within components. Indeed, this assumption defines the penalized log-likelihood function as a sum of independent functions which only depends on the partition and on the basis coefficients related to a  single dimension. Thus, its optimization can be achieved by $2J$ independent optimizations.  To obtain the triplet $\{\bOm,\Gamma,\bth\}$ maximizing the penalized observed-data log-likelihood, for a fixed number of components, many random initializations of this algorithm should be performed. Hence, the pair $\{\bmm,\bth\}$ maximizing the penalized observed-data log-likelihood is obtained by performing this algorithm for every value of $K$ between one and $K_{\max}$ and every values of $L$ between one and $L_{\max}$.

\section{Numerical experiments}\label{sec:simulation}

 Data are generated from the model defined in Section~\ref{sec:method}. Thus, each sample is composed of $n$ independent observations where each observation $\bX_{i}$ is a $J$-variate functional data observed for a period length $T_i$. The simulation setup allows different levels of dependency between the partitions as well as different numbers of irrelevant dimensions to be considered. Thus, it permits the benefits of simultaneously estimating both partitions to be investigated as well as to be performed dimension selection. 
 
 We consider two partitions with three clusters each, such that for any $k \in \{1,2,3\}$ and $\ell \in\{1,2,3\}$
$$
\mathbb{P}(V_{ik}W_{i\ell}=1) = \begin{cases}
1/3 - 2r & \text{if } k=\ell \\
r & \text{if } k\neq \ell
\end{cases},
$$
where $0\leq r \leq 1/9$ is a tuning parameter that defines the dependency between $\bV_i$ and $\bW_i$ such that $r=1/9$ leads to independence between $\bV_i$ and $\bW_i$ and $r=0$ leads to $\bV_i=\bW_i$. The observed data are functional data of dimension $J=3+s$ where the first three dimensions are relevant for estimating $\bV_i$ and $\bW_i$ and where the last $s$ dimensions are irrelevant with regard to estimating $\bV_i$ and $\bW_i$. For $j\in\{1,2,3\}$, the conditional distribution of $Z_{ij}$ given $W_{i\ell}=1$  and the conditional distribution of $Y_{ij}$ given $V_{ik}=1$ are isotropic multivariate Gaussian distributions such that $Z_{ij} \mid W_{i\ell}=1 \sim \mathcal{M}(\ell \delta (\boldsymbol{1}_5 + (2,\boldsymbol{0}_4),\delta^2 \boldsymbol{I})$ and $Y_{ij} \mid V_{ik}=1 \sim \mathcal{M}(k \delta\boldsymbol{1},\delta^2 \boldsymbol{I})$, where $\delta=0.1/\sqrt{5/3}$,  $\boldsymbol{a}_b$ is the vector of $a$ of size $b$ and $\boldsymbol{I}$ is the identity matrix of size one. For $j\in\{4,\ldots,d\}$,  the conditional distribution of $Z_{ij}$ given $W_{i\ell}=1$  and the conditional distribution of $Y_{ij}$ given $V_{ik}=1$, are isotropic multivariate Gaussian distributions such that $Z_{ij} \mid W_{i\ell}=1 \sim \mathcal{M}((2,\boldsymbol{0}_4^\top)^\top,\delta^2 \boldsymbol{I})$ and $Y_{ij} \mid V_{ik}=1 \sim \mathcal{M}(\boldsymbol{0},\delta^2 \boldsymbol{I})$. Note that the vector $(2,\boldsymbol{0}_4^\top)^\top$ is considered for the mean of $Z_{ij}$ to ensure the positivity of $\varepsilon_{ij}^2(t)$ by using \eqref{mod2}. 
For each dimension, the distribution of $\varepsilon^2_{ij}(t)$ follows \eqref{mod2} with a Fourier basis with two degrees leading to $H_j=5$ and period 125 and where each $\xi_{ij}(t)$ follows an independent centered Gaussian distribution with variance $\delta^2 5/3$. Moreover,  the distribution of $X_{ij}(t)$ follows \eqref{mod} with a Fourier basis with two degrees leading to $G_j=5$ and period 125 and where each $\varepsilon_{ij}(t)$ is obtained by the product of the square root of $\varepsilon^2_{ij}(t)$ previously generated and a Rademacher random variable. Each sequence is observed during 20 periods leading to $T_i=20 \times 125$.

To illustrate the benefits of simultaneously estimating the two partitions as well as performing dimension selection, we compare four methods according to the accuracy of their resulting partitions that is measured with the Adjusted Rand Index (ARI; \citet{hubert1985comparing}). Results of the proposed method are indicated with \emph{simult.selecTRUE} and are compared to the same approach considering that all the dimensions are relevant leading to $\Omega=\{1,\ldots,J\}$(\emph{simult.selectFALSE}), an approach assuming independence between partitions and performing a dimension selection (\emph{indpt.selectTRUE}) and an approach assuming independence between partitions and considering that all the dimensions are relevant (\emph{indpt.selectFALSE}).

\begin{figure}[ht!p]
    \centering
    \includegraphics[scale=0.4]{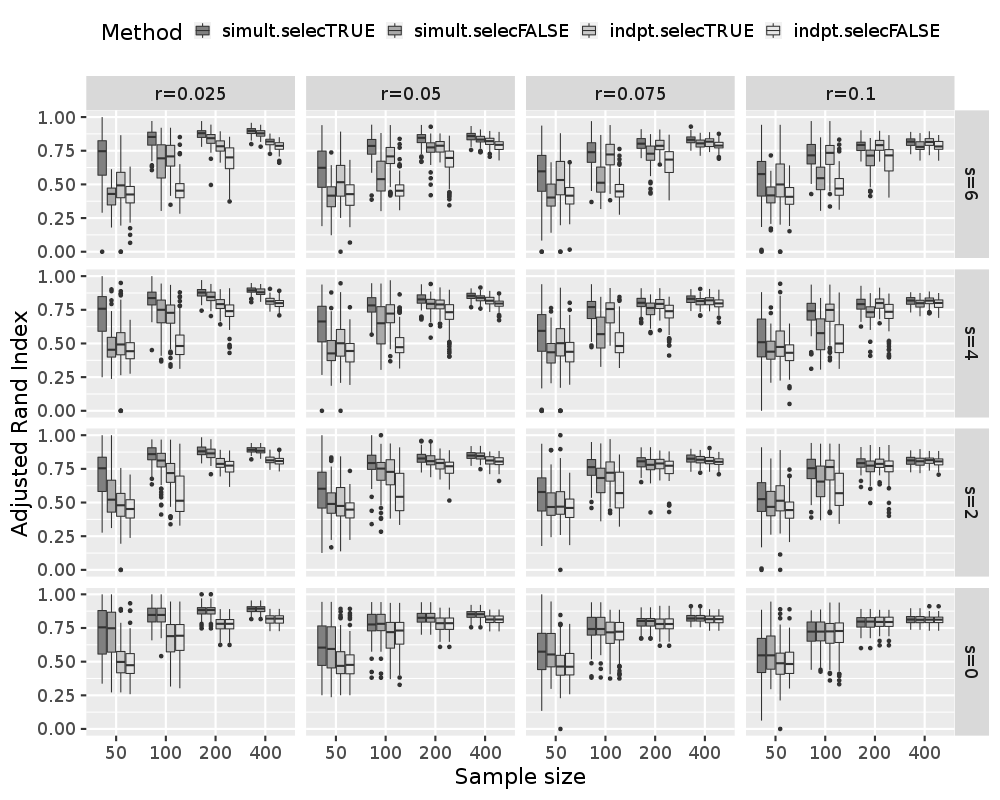}
    \caption{Boxplots of the Adjusted Rand Index obtained, for the estimation of $\bV$ depending on the number  of irrelevant variables $s$ and the dependency between the two partitions $r$, by the proposed  bi-partition clustering method with dimension selection (simult.selecTRUE), the proposed  bi-partition clustering method  without dimension selection (simult.selecFALSE), a Gaussian mixture with dimension selection  fitted only on the $\bY$  (indpt.selecTRUE) and by a Gaussian mixture without dimension selection  fitted only on the $\bY$ (indpt.selecFALSE)}
    \label{fig:partY}
\end{figure}

\begin{figure}[ht!p]
    \centering
    \includegraphics[scale=0.4]{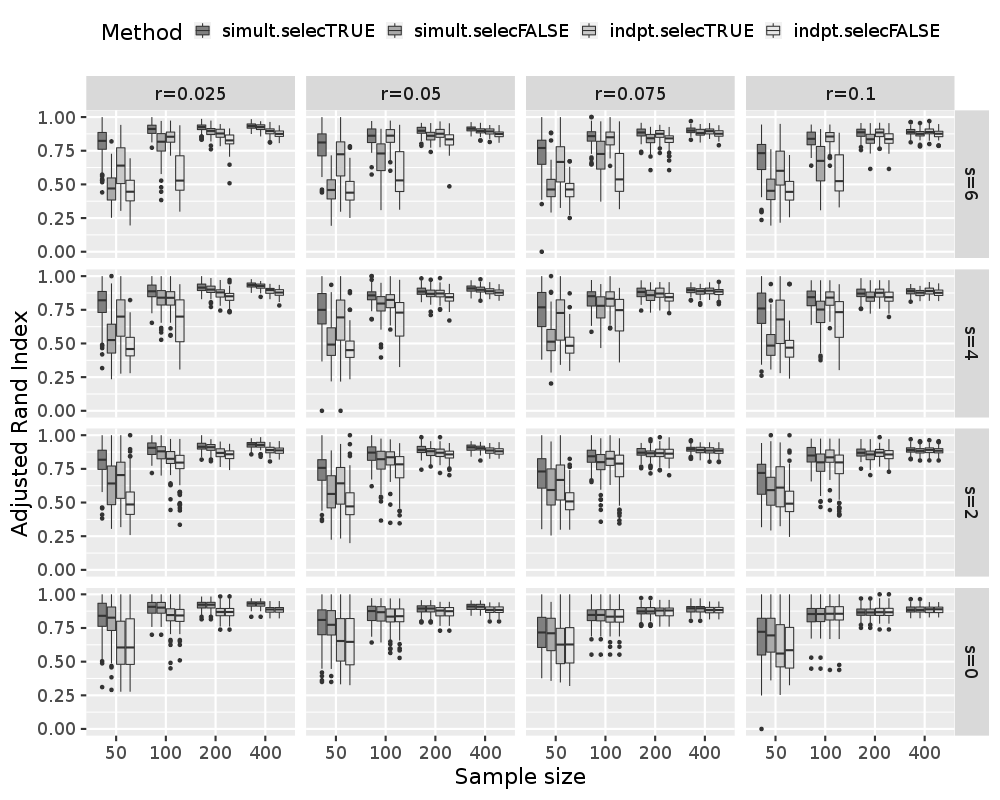}
    \caption{Boxplots of the Adjusted Rand Index obtained, for the estimation of $\bW$ depending on the number of irrelevant variables $s$ and the dependency between the two partitions $r$, by the proposed method bi-partition clustering with dimension selection (simult.selecTRUE), the proposed method without dimension selection (simult.selecFALSE), a Gaussian mixture with dimension selection  fitted only on the $\bZ$  (indpt.selecTRUE) and by a Gaussian mixture without dimension selection  fitted only on the $\bZ$ (indpt.selecFALSE)}
    \label{fig:partZ}
\end{figure}

Different situations are considered including different dependency levels between the partitions ($r=\{0,025, 0.05, 0.075, 0.1$), different numbers of irrelevant dimensions ($s=0,2,4,6$) and different sample sizes ($n=50$, 100, 200 and 400). For each situation, 100 replications are performed and the basis coefficients are estimated by considering the ordinary least squares method. Figure~\ref{fig:partY} and Figure~\ref{fig:partZ} show the boxplot of the ARI obtained by the different approaches for estimating $\bV_i$ and $\bW_i$ respectively. 

Results show that when the dependency between the two partitions are weak (see columns $r=0.1$), the methods considering the dependency obtain the same results as the methods assuming independence between partitions. However, when the dependency between the two partitions increases, it is necessary to consider this dependency. Indeed, even for the large sample, we observe that the ARI is better for the proposed method than for \emph{indpt.selecTRUE} and \emph{indpdt.selectFALSE}. In addition, results show that when all the dimension are relevant, then the proposed method obtains similar results as the method assuming that all the dimension are relevant. Moreover, when the number of irrelevant dimensions increases, performing dimension selection improves the results. This improvement is more significant if the number of irrelevant dimensions is large or if the sample size is small. As a conclusion of this experiment, it seems to be relevant to use the proposed method for any situations. Indeed, if the partitions are independent and if all the dimension are relevant for clustering, then the proposed method obtains similar results to the classic approaches. However, when there exist  dependencies between the partitions or irrelevant dimensions, then the proposed method improves the accuracy of the estimated partitions.

\section{Analysis of SWIMU data}\label{sec:appli}
This section presents the results of the proposed method for the analysis of the SWIMU data described in Section~\ref{sec:data}. 

\begin{figure}[ht!p]
    \centering
    \includegraphics[scale=0.3]{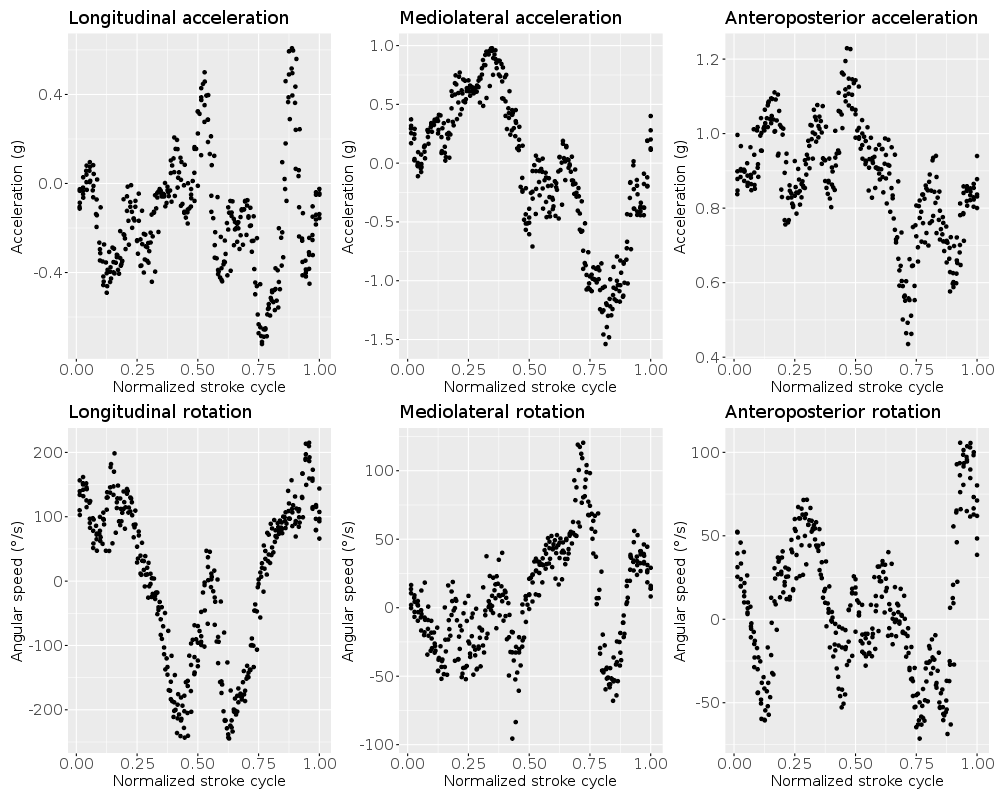}
    \caption{Example of pre-processed SWIMU data}
    \label{plotmethodo2}
\end{figure}

To deal with the issue of curve alignment, data pre-processing is conducted in order to set the first frame at the beginning of a stroke. This is achieved by using a first zero-crossing of the mediolateral acceleration after a second order Butterworth band-pass filter between 0.1 and 1Hz \citep{ganzevles2019}. 
 Since the IMU swimming records are periodic data (see Figure \ref{plotmethodo}), we decompose these multivariate functional data into a Fourier basis. The period is identified using the previously described zero-crossing that enables the stroke beginning and end to be found as show in Figure \ref{plotmethodo2}. To select the degree of the Fourier basis used for the decompositions \eqref{mod} and \eqref{mod2} of each dimension, we select the degree that minimizes the least squares error obtained by leave-one-out cross-validation. Thus, the selected degrees are between 12 and 18 for the original signal decomposition and between 8 and 12 for the decomposition of the squares residuals. The proposed methodology requires parametric assumptions about the coefficients of the functional basis. 
 
 In this application, we consider a Gaussian distribution within components with a diagonal matrix \emph{i.e.,} $ \phi_j$ is the pdf of Gaussian distribution with mean $\bmu_{jk}=(\mu_{jk1},\ldots,\mu_{jk{G_j}})^\top$ and a diagonal covariance matrix. Similarly, $\varphi_j$ is the pdf of Gaussian distribution with mean $\bnu_{j\ell}=(\nu_{j\ell1},\ldots,\nu_{j\ell{H_j}})^\top$ and a diagonal covariance matrix. The proposed methodology  permits dimension selection for a fixed number of components $K$ and $L$, depending on the BIC. Thus, the EM algorithm presented in Section~\ref{sec:Estim} is run for any $(K,L)\in\{1,\ldots,5\}^2$ and we consider the best model according to the BIC. This model is composed of 2 clusters for the swimming pattern (\emph{e.g.,} $K=2$), 3 clusters for the ability of reproducing the pattern (\emph{e.g.,} $L=3$). Moreover, five among the six dimensions are selected for the swimming pattern (all the dimensions but the mediolateral rotation) and all the dimensions are selected for the ability of reproducing the pattern. For the selected model, the assumption of Gaussian components has been investigated using a Shapiro-Wilk normality test for each coefficient, by considering a Bonferroni correction to circumvent the issue of multiple testing procedures. Considering a nominal level of 0.05, less than $3\%$ of the test statistics belong to the rejection region and thus, the normality assumption seems to be reasonable for analyzing SWIMU data. The assumption of independence of the coefficients within a component is also investigated (\emph{i.e.,} decomposition \eqref{eq:decompogr} as well as considering diagonal covariance matrices for the Gaussian densities $ \phi_j$ and $\varphi_j$). By considering   Pearson's product-moment correlation with a  Bonferonni correction and a nominal level of 0.05 means that less than $1\%$ of the test statistics belong to the rejecting region and hence, this assumption seems to be relevant for analyzing the SWIMU data.

\begin{figure}[ht!p]
    \centering
    \includegraphics[scale=0.35]{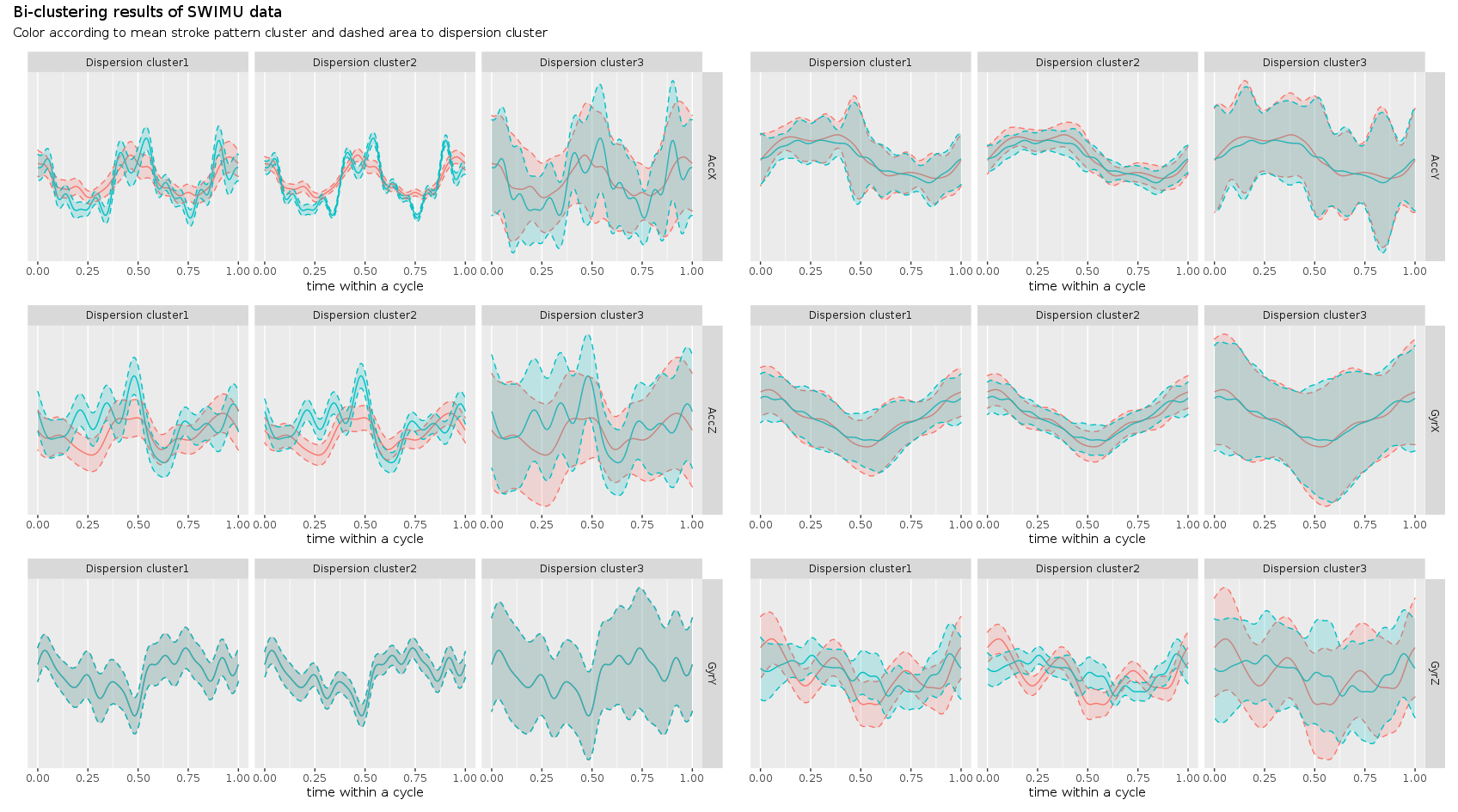}
    \caption{Classes of swimming pattern (\emph{i.e.,} mean curves $\overline{X}_{jk}(t)$ in plain) and classes of abilities to reproduce the swimming pattern (\emph{i.e.,} credibility regions $[\overline{X}_{jk}(t) -2\overline{\varepsilon}_{j\ell}(t),\overline{X}_{jk}(t) +2\overline{\varepsilon}_{j\ell}(t)]$ defined by the dashed lines)}
    \label{plotbiclustering}
\end{figure}

Figure~\ref{plotbiclustering} allows for an easy interpretation of the double partition obtained on the swimming data. Indeed, it summarizes the swimming patterns by presenting in plain lines, for each dimension $j$, the mean curves defined, for any $k\in\{1,2\}$ and $t\in[0,1[$, by $\overline{X}_{jk}(t) :=   \widehat{\bmu}_{jk}^\top  \bpsi_{j}(t;1)$. Moreover, the three clusters of abilities for reproducing the swimming pattern are summarized, for each dimension $j$, by the credibility region, at each time $t$, $[\overline{X}_{jk}(t) -2\overline{\varepsilon}_{j\ell}(t),\overline{X}_{jk}(t) +2\overline{\varepsilon}_{j\ell}(t)]$  defined as the area that differs from two standard deviations at each time $t$ from the mean curve where, for any $\ell\in\{1,2,3\}$ and $t\in[0,1[$,
$\overline{\varepsilon}^2_{j\ell}(t) :=   \widehat{\bnu}_{j\ell}^\top  \bPi_{j}(t;1)$. 
Figure~\ref{plotbiclustering} shows that dimension selection procedure excludes the mediolateral rotation as an informative feature only for the clustering of the swimming pattern. The related red and blue curves are overlap  but the different credibility regions do not. Mediolateral rotation mainly reflects the pitch movement of the swimmers and so is not a discriminant constitutive motion of technical abilities for front-crawl sprint.

 The partition of swimming patterns is composed of a majority class (81\%) presenting a fluent acceleration pattern with continuity of propulsive actions during the stroke cycle. This class, called the \textit{smoothy} class (see red curves in Figure~\ref{plotbiclustering}), is characterized by a lower acceleration variation on the propulsive axis (\emph{i.e.,} longitudinal and anteroposterior acceleration) and a higher rotation variation on both longitudinal and anteroposterior axes in addition to mediolateral acceleration.  The second class of swimming patterns (19\%) is composed of more explosive and irregular stroke patterns. It includes higher acceleration variations than in the \textit{smoothy} class particularly on the propulsive axis, associated with reduced mediolateral acceleration and rotation variations inducing more stability and a better alignment of the body. This steadiness especially occurs during the critical propelling phases of the cycle (\emph{i.e.,} peaks on longitudinal acceleration).  This class is then called the \textit{jerky} class (see blue curves in Figure~\ref{plotbiclustering}).
 
The partition of abilities for reproducing the swimming patterns is composed of two classes with equal proportions (37\%) characterized by moderate and high stroke pattern repeatability, and of a third class (26\%) with low level of stroke repeatability. We respectively call them the \textit{moderate}, \textit{high} and \textit{low repeatability} classes. It is interesting to note that the credibility region around mean curves are not constant during the swimming pattern indicating unequal variance. In this way, the functional visualisation of stroke pattern dispersion allowed by the model provides a useful visualisation of kinematical variability during some specific and determinant stroke cycle phases. 
  
The model allows us to take into account the dependencies between partitions. Modeling this dependency between the two partitions is crucial since the difficulty of reproducing a swimming pattern depends on its kinematics. Table \ref{crosspartition} presents the contingency table between the two partitions. Note that the assumption of independence between both partitions is rejected according to a Pearson's Chi-squared test  ($\chi^2$=15.62, df=2, $p$<0.01). This confirms the biomechanical association between the swimming pattern and the ability to reproduce it as a discriminant feature of technical skills for front-crawl sprint swimming.  \\

\begin{table}[ht!p]
\centering
\begin{tabular}{rcc}
  \hline
 & \textit{Jerky} & \textit{Smoothy} \\ 
  \hline 
\textit{High repeatability} &   1 &  24 \\ 
\textit{Moderate repeatability} &   3 &  22 \\ 
\textit{Low repeatability} &   9 &   9 \\ 
   \hline
\end{tabular}
    \caption{Contingency table between the two partitions: swimming pattern (by column) and abilities to reproduce the pattern (by row).}
    \label{crosspartition}
\end{table}
The two partitions clustering allows technical skills to be measured. This interpretation can be confirmed by biomechanical parameters. Thus, Table~\ref{ParamPartition1} presents statistics related to the interpretation of the classes of swimming pattern. Indeed, it presents the standard deviations of the mean curves, defined as $[\int_0^1 (\overline{X}_{jk}(t) - \int_0^1 \overline{X}_{jk}(t')dt')^2dt]^{1/2}$ that reflect cycle variation and the jerk cost  that is a practical evidence of stroke smoothness (\cite{ganzevles2019}; \cite{dadashi2016front}), defined as $ (\overline{X}_{jk})=[\int_0^1 (\overline{X}_{jk}'(t))^2dt]$ where $\overline{X}_{jk}'(t)$ stands for the derivative of the mean curve $\overline{X}_{jk}(t)$. Note that jerk cost can be computed only for the dimensions related to acceleration. These indicators confirm the visual observations about swimming patterns with about a twice higher longitudinal acceleration standard deviation and a two thirds lower anteroposterior rotation standard deviation for \textit{jerky} class comparing to \textit{smoothy} class. 
\begin{table}[ht!p]
\centering
\begin{tabular}{rrccccc}
  \hline
  & Cluster & \multicolumn{3}{c}{Acceleration}& \multicolumn{2}{c}{Rotation} \\ 
  \
  &   & Longitudinal   & Mediolateral   & Anteroposterior   & Longitudinal   & Anteroposterior   \\ 
  \hline
Standard deviation & \textit{Jerky} & 0.16 & 0.52 & 0.08 & 108.65 & 12.56 \\ 
  & \textit{Smoothy} & 0.09 & 0.57 & 0.04 & 113.83 & 16.38 \\ 
  \hline
Jerk cost & \textit{Jerky} & 2.84x$10^5$ & 2.38x$10^5$ & 0.71x$10^5$ &   &   \\ 
  & \textit{Smoothy} & 0.19x$10^5$ & 1.93x$10^5$ & 0.04x$10^5$ &   &   \\ 
   \hline
\end{tabular}
\caption{Standard deviation and jerk cost by dimension of the mean curve of the classes of    patterns of swimming.}
\label{ParamPartition1}
\end{table} 

Table~\ref{ParamPartition2} presents the standard deviations of the mean curves $[\int_0^1 (\overline{Y}_{j\ell}(t) - \int_0^1 \overline{Y}_{j\ell}(t')dt')^2dt]^{1/2}$. Regarding ability to reproduce the swimming pattern, the longitudinal acceleration standard deviation is for example more than twice higher in \textit{moderate repeatability} than the \textit{high repeatability} class and four times higher in \textit{low repeatability} than the \textit{moderate repeatability} class.

\begin{table}[ht!p]
\centering
\begin{tabular}{rcccccc}
  \hline
   Cluster & \multicolumn{3}{c}{Acceleration}& \multicolumn{2}{c}{Rotation} \\ 
  \
     & Longitudinal   & Mediolateral   & Anteroposterior   & Longitudinal   & Mediolateral   &Anteroposterior   \\ 
  \hline
\textit{High repeatability} & 1.72x$10^3$ & 3.25x$10^3$ & 0.74x$10^3$ & 100 & 31.0 & 33.2 \\ 
\textit{Moderated repeatability} & 3.62x$10^3$ & 10.2x$10^3$ & 1.24x$10^3$ & 250 & 47.8 & 57.2 \\ 
\textit{Low repeatability} & 14.5x$10^3$ & 13.4x$10^3$ & 4.77x$10^3$ & 503 & 175.0 & 66.6 \\ 
   \hline
\end{tabular}
\caption{Standard deviation by dimension of the mean curve of the classes of abilities of reproducing the swimming pattern.}
\label{ParamPartition2}
\end{table}

\begin{figure}[ht!p]
    \centering
    \includegraphics[scale=0.3]{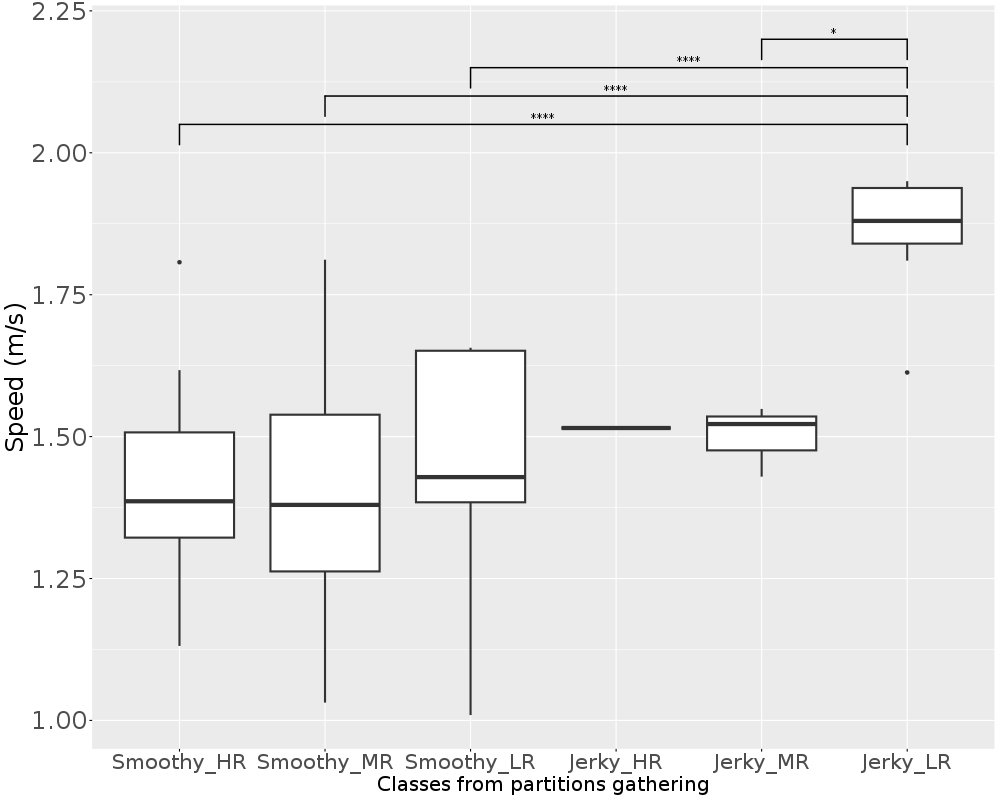}
    \caption{Boxplot of sport performance (i.e., swimming speed) according to gathered partitions from clustering defining swimming pattern and ability to reproduce it. Stars indicate significant pairwise comparisons between classes: * p<0.05, **** p<0.0001}
    \label{plotperf}
\end{figure}

We now investigate the relation between the technical skills reflected by estimated partitions and performance.  Figure \ref{plotperf} illustrates this relation by presenting a boxplot of the swimming speed according to gathered clusters. There is a significant effect of gathered partitions on swimming speed as shown by a Fisher test (F(5.62)=10.1, p<0.001, $\eta^2$=0.45). The \textit{jerky} swimming pattern associated with \textit{low repeatability} is the fastest biomechanical strategy (1.86±0.10 m/s). Indeed, all its pairwise comparisons with others clusters are significant considering a nominal level of 0.05 (except with the \textit{jerky}+\textit{high repeatability} class but there is only one swimmer in this cluster). There are no other significant pairwise comparisons between all kinds of other clusters. \\
Our approach allows us to discriminate the performance level of the swimmers since there is a clear speed trend for the \textit{jerky}+\textit{low repeatability} cluster. Thus, it enables us to group swimmers of homogeneous technical skills regarding performance. Furthermore this approach allows us to perform technical skills evaluation according to stroke cycle kinematics functional modelling on the micro-scale, instead of the classic macro-scale spatio-temporal parameters such as stroke rate, stroke length and stroke index. In this way, our double partition clustering provides a valuable sports-related outcome.

\section{Conclusion} \label{sec:conclusions}
We have developed a model-based approach that provides two complementary partitions allowing technical skills to be tracked in swimming based on multivariate functional data. The model considers both the kinematical information from the original signals (\emph{i.e.,} measuring the swimming pattern) and the residuals resulting from the functional basis decomposition (\emph{i.e.,} measuring the abilities to reproduce the swimming pattern) and allows for dependency of the two partitions. Otherwise, the method allows for dimension selection to better establish the biomechanical contribution to technical skills.\\
The results of the application confirm the double partition model's sensitivity to aggregate kinematical variability defining swimming technical skills. Clustering IMU sprint front crawl data highlights specific kinds of stroke kinematics related to speed production. In this way it allows us to identify a relevant biomechanical strategy for front-crawl sprint performance relying on a jerky unstable swimming pattern. Technical skills are driven by management of kinematical variability through, on the one hand, specific swimming patterns linked to stroke smoothness defining continuity of propulsive actions and energy expenditure, and on the other hand, repeatability defining pattern stability and ability to reproduce swimming strokes.\\
The development of this procedure expands traditional technique monitoring by avoiding only relying on human-based observations of experts or on macro-scale spatio-temporal parameters recorded by a stopwatch. Hence it gives to coaches a complementary data-driven method less subjective to current eyes-based method. Firstly, to better establish skills evaluation in environmental conditions of daily training and secondly to better characterize technical levels of front-crawl swimmers through an automatic user-friendly framework. The proposed model could be applied to a wide population of athletes with different characteristics and leads to biomechanical profiling of swimmers. It provides the first in-situ modelling of swimming IMU data in the literature.\\ 
An interesting perspective is to consider extending of this approach to longer distances including fatigue and degradation of the stroke. These characteristics of swimming could impact the IMU signals resulting in more complex methodology and challenging the modelling of functional data. In particular, because fatigue induced by these exercises, deteriorates technical skills. As a consequence, it implies that biomechanical regulations of stroke variability leading to non-constant periodicity  of swimming patterns that produce non-stationary multivariate time series as well as a deterioration of the abilities to reproduce the swimming patterns .

\begin{appendix}
%Section~\ref{supp:maths} presents the proofs and the %mathematical details about the results stated in the paper. 
%Section~\ref{supp:appli} presents supplementary information %about the analysis of the IMU data.
%Section~\ref{supp:assumption} presents supplementary figures %about models assumptions.

\section{Proofs}\label{supp:maths}

\begin{proof}[Proof of Lemma~\ref{lem:idengrl}]
The parameters of model \eqref{eq:modelgrl} are identifiable when we have for any $\btheta\in\Theta$ and $\widetilde{\btheta}\in\Theta$
$$
\forall (\by^\top,\bz^\top)^\top,\; p(\by,\bz;\btheta)= p(\by,\bz;\widetilde{\btheta}) \Longrightarrow \btheta = \widetilde{\btheta}.
$$
The marginal distributions of $\bY$ and $\bZ$ are defined by the following pdf
$$
p(\by;\brho,\balpha) = \int p(\by,\bz;\btheta) d\bz = \sum_{k=1}^K \rho_k f_k(\by;\balpha_k),
$$
and
$$
p(\bz;\bnu,\bbeta) = \int p(\by,\bz;\btheta) d\by = \sum_{\ell =1}^L \nu_\ell g_\ell(\bz;\bbeta_\ell),
$$
where $\brho=(\rho_1,\ldots,\rho_K)^\top$, $\balpha=(\balpha_1^\top,\ldots,\balpha_K^\top)^\top$, $\bnu=(\nu_1,\ldots,\nu_L)^\top$, $\bbeta=(\bbeta_1^\top,\ldots,\bbeta_L^\top)^\top$.
Let $\btheta$ and $\widetilde{\btheta}$ be two parameters such that
\begin{equation}\label{eq:startID}
\forall (\by^\top,\bz^\top)^\top,\; p(\by,\bz;\btheta)= p(\by,\bz;\widetilde{\btheta}).
\end{equation}
Since the parameters of the marginal distributions of $\bY$ and $\bZ$ are identifiable, then \eqref{eq:startID} implies that 
\begin{equation}\label{eq:equality}
\balpha = \widetilde{\balpha},\;\brho = \widetilde{\brho},\;\bbeta = \widetilde{\bbeta} \text{ and } \bnu = \widetilde{\bnu}.
\end{equation}
To conclude the proof, we need to show that  \eqref{eq:startID} implies that $\pi_{k\ell}=\widetilde{\pi}_{k\ell}$. A direct consequence of the marginal identifiability of the parameters of the marginal distributions is that the pdf $f_1,\ldots,f_K$ are linearly independent as well as the pdf  $g_1,\ldots,g_L$. Therefore, for any $\bu\in\mathbb{R}^K$ and $\bv\in\mathbb{R}^L$, we have
\begin{equation}\label{eq:line1}\forall \by,\; \sum_{k=1}^K u_k f_k(\by;\balpha)=0  \Longrightarrow \bu = \boldsymbol{0}_K\end{equation} 
and
\begin{equation}\label{eq:line2}\forall \bz,\; \sum_{\ell=1}^L v_\ell g_\ell(\bz;\bbeta)=0  \Longrightarrow \bv = \boldsymbol{0}_L,\end{equation} 
$\boldsymbol{0}_d$ being the vector of length $d$ composed of zeros. Using the equality constraints \eqref{eq:equality}-\eqref{eq:startID} implies that
$$
\forall (\by^\top,\bz^\top)^\top,\;  \sum_{k=1}^K \sum_{\ell=1}^L (\pi_{k\ell} - \widetilde{\pi}_{k\ell})f_k(\by;\balpha_k)g_\ell(\bz;\bbeta_\ell))=0.
$$
Denoting by $\pi_{k\mid \ell}$ the conditional probability that $V_k=1$ given $W_\ell=1$ and noting that \eqref{eq:equality} implies $\nu_{\ell}=\widetilde{\nu}_{\ell}$, the previous equation being equivalent to
$$
\forall (\by^\top,\bz^\top)^\top,\;  \sum_{k=1}^K \left[\sum_{\ell=1}^L (\pi_{k\mid \ell} - \widetilde{\pi}_{k\mid \ell}) \rho_{\ell} g_\ell(\bz;\bbeta_\ell) \right] f_k(\by;\balpha_k)=0.
$$
Using \eqref{eq:line1} with $u_k=\sum_{\ell=1}^L (\pi_{k\mid \ell} - \widetilde{\pi}_{k\mid \ell}) \rho_{\ell} g_\ell(\bz;\bbeta_\ell) $, we have that \eqref{eq:startID} implies that
$$
\forall \bz,\; \sum_{\ell=1}^L (\pi_{k\mid \ell} - \widetilde{\pi}_{k\mid \ell}) \rho_{\ell} g_\ell(\bz;\bbeta_\ell) = 0.
$$
Using \eqref{eq:line2} with $v_{\ell} =  (\pi_{k\mid \ell} - \widetilde{\pi}_{k\mid \ell}) \rho_{\ell}$ and noting that identifiability of the parameters of the marginal distribution of $\bZ$ implies that $\rho_{\ell}>0$. We have that \eqref{eq:startID} implies $\pi_{k\mid \ell } = \widetilde{\pi}_{k\mid \ell}$ and so $\pi_{k\ell} = \widetilde{\pi}_{k\ell}$. Combining $\pi_{k\ell } = \widetilde{\pi}_{k\ell}$ and \eqref{eq:equality} leads to the identifiability of the parameters of model \eqref{eq:modelgrl}.

\end{proof}

\begin{proof}[Proof of Lemma~\ref{lem:partitionaccuracy}]
We show the results for the classification rules related to $\bV_i$. The same reasoning can be applied to state the results related to $\bW_i$. The posterior probabilities of classification, for the estimation of $\bV_i$, based on $(\bY_i^\top,\bZ_i^\top)^\top$ and $\bY_i$ are respectively defined by
$$
t_k(\by_i,\bz_i):=\mathbb{P}(V_{ik}=1\mid \bY_i=\by_i,\bZ_i=\bz_i) = \frac{\pi_{k\bullet} f_k(\by_i)h_k(\bz_i)}{p(\by_i,\bz_i;m,\btheta)},
$$
and
$$t_k(\by_i):=\mathbb{P}(V_{ik}=1\mid \bY_i=\by_i) = \frac{\pi_{k\bullet} f_k(\by_i)}{\sum_{k'=1}^K\pi_{k'\bullet} f_{k'}(\by_i)},
$$
where $\pi_{k\bullet}=\sum_{\ell=1}^L\pi_{k\ell}$, $h_k(\bz_i)=\sum_{\ell=1}^L \rho_{\ell \mid k} g_\ell(\bz_i)$ and $\rho_{\ell\mid k} = \pi_{k\ell}/\pi_{k\bullet}$. Let the classification regions be defined by assigning an observation to the most likely cluster as follows
$$
R_k^{Y,Z} = \{(\by^\top,\bz^\top)^\top: \forall k'\neq k\; t_k(\by,\bz)>t_{k'}(\by,\bz) \}
\text{ and }
R_k^{Y} = \{\by: \forall k'\neq k\; t_k(\by)>t_{k'}(\by) \}.
$$
Note that by definition the classification rule defined by assigning an observation to the most likely cluster based on $t_k(\by,\bz)$, is optimal. Hence, it suffices to show that  $\Upsilon_V(\bY,\bZ)$ and $\Upsilon_V(\bY,\bZ)$ defined in Lemma \ref{lem:partitionaccuracy} are not equal on a set of non-zero null Lebesgue measure. 

Since $\bV_i$ and $\bW_i$ are not independent, there exist $k_1$ and $k_2$ and $\tilde\ell$ such that $\rho_{\tilde \ell \mid k_1}\neq\rho_{\tilde \ell \mid k_2}$. By the identifiability of the marginal distribution of $\bZ_i$, the densities $g_1,\ldots,g_L$ are not linearly independent meaning there exists $\tilde\bz$ such that $\sum_{\ell=1}^L \left(\rho_{\tilde \ell \mid k_2}-\rho_{\tilde \ell \mid k_1}\right) g_{\ell}(\tilde \bz)\neq 0$. Thus, there exists $\varepsilon>0$ and $\rho_\varepsilon>0$ such that for any $\bz\in \mathcal{V}_{\rho_\varepsilon}(\tilde \bz)=\{\bz: \|\bz - \tilde\bz\| <\rho_{\varepsilon}\}$, $h_{k_2}(\bz)/h_{k_1}(\bz)>1+\varepsilon$. Let $S_{k_1,k_2,\varepsilon}^Y$ be the subset of $R_{k_1}^Y$ defined by
$$
S_{k_1,k_2,\varepsilon}^Y = \{ \by \in R_{k_1}^Y : \frac{t_{k_1}(\by)}{t_{k_2}(\by)} < 1 + \varepsilon
\}.
$$
By definition, $\forall \by \in S_{k_1,k_2,\varepsilon}^Y$, $\Upsilon_V(\by)$ affects the observation into cluster $k_1$. Moreover, $\forall (\by^\top,\bz^\top)^\top \in S_{k_1,k_2,\varepsilon}^Y \times \mathcal{V}_{\rho_\varepsilon}(\tilde \bz)$, we have $\pi_{k_2} f_{k_2}(\by) h_{k_2}(\bz)>\pi_{k_1} f_{k_1}(\by) h_{k_1}(\bz)$ and thus $\Upsilon_V(\by,\bz)$ does not affect the observation into cluster $k_2$. The proof is concluded by noting that $S_{k_1,k_2,\varepsilon}^Y \times \mathcal{V}_{\rho_\varepsilon}(\tilde \bz)$ does not have a null Lebesgue measure.
\end{proof}

\begin{proof}[Proof of Lemma~\ref{lemma:IDfinal}]
It suffices to show that the marginal distributions of $\bY_i$ and $\bZ_i$ are identifiable. Indeed, once this is done, we can apply Lemma~\ref{lem:idengrl} to conclude the proof.
Consider $\brho$, $\balpha$, $\widetilde{\brho}$, $\widetilde{\balpha}$ such that
$$
\forall \by_i,\; p(\by_i;\brho,\balpha)= p(\by_i;\widetilde{\brho},\widetilde{\balpha}).
$$
Since, by assumption, there exists $j_0\in\Omega$ such that the parameters of the marginal distribution of $\bY_{ij}$ are identifiable, we have
$$
\forall \by_{ij_0},\; \sum_{k=1}^K \rho_k \phi_{j_0}(\by_{ij_0};\balpha_{kj_0})=\sum_{k=1}^K \widetilde{\rho}_k \phi_{j_0}(\by_{ij_0};\widetilde{\balpha}_{kj_0})
\Longrightarrow \rho_k=\widetilde{\rho}_k \text{ and } \balpha_{kj_0}=\widetilde{\balpha}_{kj_0} \text{ for any } k=1,\ldots,K.
$$
Thus, for any $\bu\in\mathbb{R}^K$ and $\bv\in\mathbb{R}^L$, we have
\begin{equation}\label{eq:singleJ}\forall \by_{ij_0},\; \sum_{k=1}^K u_k \phi_{j_0}(\by;\balpha_{j_0})=0  \Longrightarrow \bu = \boldsymbol{0}_K.\end{equation} 
We now consider the distribution of the pair $(\bY_{ij_0}^\top,\bY_{ij}^\top)^\top$, for any $j\neq j_0$. This distribution implies that
$$
\forall (\by_{ij_0}^\top,\by_{ij}^\top)^\top,\; \sum_{k=1}^K \rho_k \phi_j(\by_{ij};\balpha_{kj}) \phi_{j_0}(\by_{ij_0};\balpha_{kj_0})=\sum_{k=1}^K \rho_k \phi_{j}(\by_{ij};\widetilde{\balpha}_{kj})\phi_{j_0}(\by_{ij_0};\balpha_{kj_0}).
$$
Note that if $j\in\Omega^c$, then we have $\balpha_{1j}=\ldots=\balpha_{Kj}$. Using \eqref{eq:singleJ} with $u_k=\rho_k(\phi_j(\by_{ij};\balpha_{kj})-\phi_{j}(\by_{ij};\widetilde{\balpha}_{kj}))$ leads to $\balpha_{kj}=\widetilde{\balpha}_{kj}$ for any $k$. Using this reasoning for any $j$ leads to the identifiability of the parameters of the marginal distribution of $\bY_i$. Similar approaches can be used to state the identifiability of the parameters of the marginal distribution of $\bZ_i$.
\end{proof}
\end{appendix}

\bibliographystyle{apalike} % Style BST file
\bibliography{Biblio_StageENSAI}       % Bibliography file (usually '*.bib')

\end{document}